\documentclass[aip,rsi,reprint,graphicx]{revtex4-1} 

\usepackage{graphicx}

\begin{document}

\title{Spatial shaping for generating arbitrary optical dipoles traps for ultracold degenerate gases} 

\author{Jeffrey G. Lee}
\email[]{jglee@umd.edu}
\affiliation{Joint Quantum Institute, University of Maryland, College Park, Maryland 20742}
\affiliation{Institute for Physical Science and Technology, University of Maryland, College Park, Maryland 20742}
\author{W.~T.~Hill, III}
\email[]{wth@umd.edu}
\affiliation{Joint Quantum Institute, University of Maryland, College Park, Maryland 20742}
\affiliation{Institute for Physical Science and Technology, University of Maryland, College Park, Maryland 20742}
\affiliation{Department of Physics, University of Maryland, College Park, Maryland 20742}

\date{\today}

\begin{abstract}
We present two spatial-shaping approaches -- phase and amplitude -- for creating two-dimensional optical dipole potentials for ultracold neutral atoms.  When combined with an attractive or repulsive Gaussian sheet formed by an astigmatically focused beam, atoms are trapped in three dimensions resulting in planar confinement with an arbitrary network of potentials -- a free-space atom chip.  The first approach utilizes an adaptation of the generalized phase-contrast technique to convert a phase structure embedded in a beam after traversing a phase mask, to an identical intensity profile in the image plane. Phase masks, and a requisite phase-contrast filter, can be chemically etched into optical material (e.g., fused silica) or implemented with spatial light modulators; etching provides the highest quality while spatial light modulators enable prototyping and realtime structure modification.   This approach was demonstrated on an ensemble of thermal atoms. Amplitude shaping is possible when the potential structure is made as an opaque mask in the path of a dipole trap beam, followed by imaging the shadow onto the plane of the atoms. While much more lossy, this very simple and inexpensive approach can produce dipole potentials suitable for containing degenerate gases. High-quality amplitude masks can be produced with standard photolithography techniques.  Amplitude shaping was demonstrated on a Bose-Einsten condensate.      

\end{abstract}

\pacs{}

\maketitle

\section{Introduction}
Many recent proposals and experiments in neutral atom physics have centered around two-dimensional (2D) or quasi-2D trap geometries. These sheet traps are generally created using optical dipole potentials\cite{ODP} with either tightly focused, highly elliptical Gaussian beams\cite{Lee:2013,Wright_Phase_Slip,Ramanathan_Persistent_Current}, or one-dimensional optical lattices\cite{Greiner_2009} confining atoms to a sheet. On top of these sheets, a 2D pattern of some sort is projected to create the desired 2D potential landscape. These patterns are generally created using propagating modes of a laser. One example of this is the recent use of Laguerre-Gaussian modes to trap Bose-Einstein condensates in ring geometries to study persistent currents and phase slips between quantized angular momentum states in atomic analogs to superconducting wire rings\cite{Wright_Phase_Slip,Ramanathan_Persistent_Current}. Other experiments have used simple Gaussian modes, rapidly translated using acousto-optic deflectors (AOD) to create a wide variety of potential\cite{Painted_Potentials}. Both of these methods have advantages and disadvantages. Using AODs provides an easy method of creating dynamic potentials, with time varying structure, however higher heating rates have been seen in traps that are composed exclusively of a scanned Gaussian beam\cite{AOD_Heating}. More complex modes do not share this disadvantage and are generally quite stable, but are limited in the geometries that can be realized. Here we discuss two methods of creating arbitrary 2D ODP that we have developed. Each of these overcomes the disadvantages mentioned above, while bringing up new issues that need to be taken into account when designing new experiments. Both of these would benefit from recent advances in high numerical aperture (NA) trapping and imaging systems\cite{Greiner_2009}.

\section{Phase Shaping}
\begin{figure*}
\includegraphics{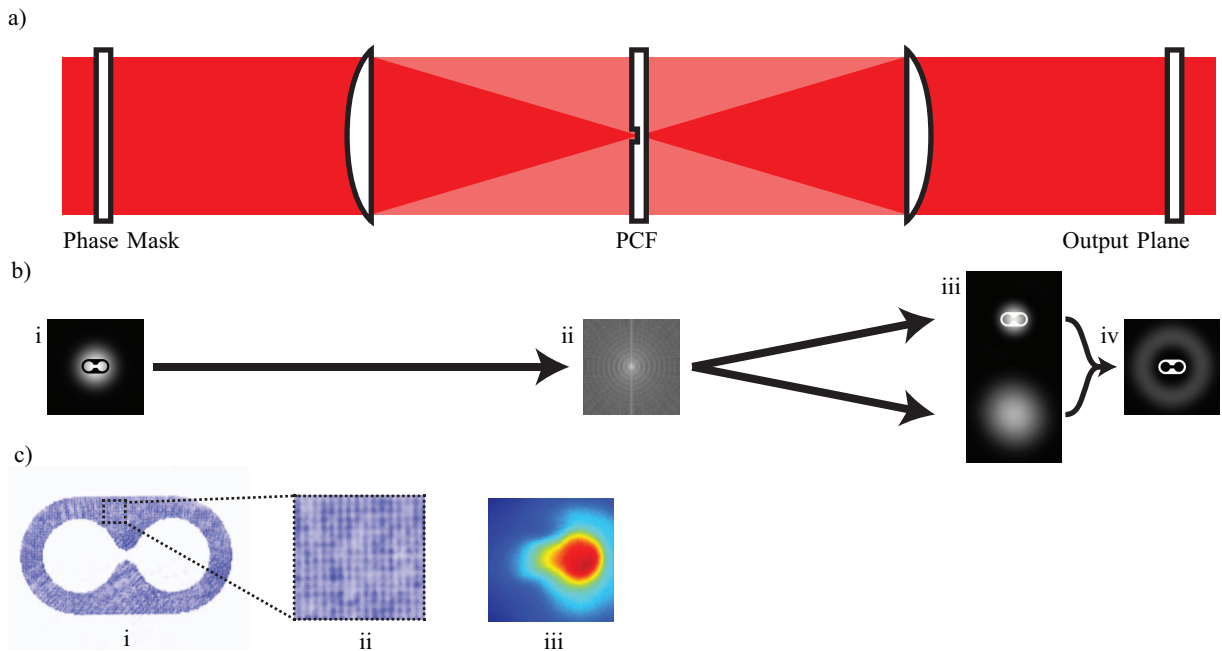}
\caption{a) Schematic of the GPC optical setup, consisting of an input phase mask, two lenses in a 4-f configuration, and a PCF. b) A representation of the dipole trap beam field as it passes through the setup taken from simulations, including i) the input beam with desired phase mask (dark area to represent phase), ii) the intensity pattern at the Fourier plane (log scale), iii) the two portions of the beam at the output plane that either did (lower) or did not (upper) pass through the PCF (light region to represent phase), and iv) the output intensity pattern resulting from the interference of the portions in iii). c) i) A typical CCD camera image at the output plane of the GPC optical setup for the example pattern from b), ii) a close-up image of this pattern, showing the dead-space artifacts from the SLM, and iii) a fluorescence image of thermal ($\sim 20 \mu$K), $^{85}$Rb atoms, taken 2~ms after release from the potential displayed in i)\cite{Lee:2013}. Note, the atoms were released and imaged while contained on only one side of the dumbbell shaped potential. \label{fig:GPC}}%
\end{figure*}

Phase shaping (PS) exploits the generalized phase-contrast (GPC) approach \cite{Gluckstad:1996, Alonzo:2007, Palima:2007}, which is an efficient method for spatially shaping the transverse intensity profile a focused laser beam.  The GPC technique redirects most of the beam intensity into the desired potential pattern, making efficient use of available laser power.  As a derivative of the Zernike phase imaging \cite{Zernike:1955}, it relies on a strategically sized and placed phase-contrast filter (PCF) to enable a one-to-one mapping of a phase embedded across a laser beam to a high-contrast spatial intensity profile. While we only show this for a binary phase pattern, the theory described by Gluckstad et al. extends fully to continuous phase patterns. At the same time, this makes phase-contrast shaping closely related to phase-contrast imaging that has been used to monitor atomic cloud densities in a minimally destructive way \cite{Andrews:1997}.  We demonstrate the approach with the setup shown in Figure~\ref{fig:GPC}, which is easily integrated into standard imaging systems used to study cold atoms.  While we employed a 4-f arrangement to create the intensity pattern in our demonstration of the technique\cite{Lee:2013}, other arrangements are also possible.  The GPC approach requires the phase mask to be placed in a collimated beam before the first lens and the PCF to be located in the Fourier plane of the first lens.  The desired intensity pattern appears in the image plane of the second as shown in Figure~\ref{fig:GPC}.  Typically, the focal lengths needed for best application of the GPC approach are not commensurate with the required image size and the location for the atoms.  Thus, the image plane after the second lens is relayed, and potentially resized, by a second pair of lens to the atoms inside the vacuum chamber.  

In our demonstration of this technique, we used a computer controlled spatial light modulator (SLM) as the phase mask\cite{Lee:2013}. Etched phase masks could also be used for this purpose. While they will have better resolution, they lack the flexibility that SLMs offer for real time changes to the masks. The PCF is a simple phase mask consisting of a circular region that shifts the phase of low spatial frequencies by $\pi$. The diameter of the circle depends on the input beam size and the details of the phase mask. The initial PCF diameter is chosen to match the focal spot size of the beam in the Fourier plane, with the phase mask removed. Typically the diameter has to be adjusted to optimize the setup for maximum contrast as discussed below. 

The PCF splits the beam into two parts. One part contains the lowest-order spatial frequencies that have been shifted in phase by $\pi$. The profile of this beam will be close to Gaussian because it is nearly devoid of any higher order spatial frequencies. The phase of the other part is unshifted. This portion contains the initial Gaussian envelope, in addition to all of the high spatial frequency components of the associated phase pattern. The two parts will interfere in the image plane to produce the desired intensity profile. The sequence of events is depicted in Figure~\ref{fig:GPC}. For optimum efficiency and contrast, the intensities of the two parts of the beam should be equal, placing restrictions of the size of the PCF and the particular phase pattern used as mentioned above. Due to mismatch in the sizes of the two parts, there is typically a ring of light surrounding the desired pattern. If problematic, this can be blocked with an iris before subsequent imaging onto the atoms. In practice, it is useful to simulate desired patterns using numerical approximations.

In practice, due to having a limited number of PCFs available, and the effort required to change the beam size, we typically fix the PCF and beam sizes, and vary the size of the phase pattern to maximize the contrast in the output plane. For our demonstration\cite{Lee:2013}, the $1/e^2$ radius of the initial beam was 10~mm, and the focal length of the lenses were 60~mm. The SLM (Hamamatsu PAL-SLM) used to produce the initial phase pattern was a 768-by-768 array, covering an area of 2.5-by-2.5~mm. The PCF was created in collaboration with the Laboratory for Physical Sciences, and consists of a fused silica window with an array of nine phase spots on a 5~mm grid, varying from 5.5 to 6.5~$\mu$m. The phase spots were chemically etched to a depth of 850~nm, producing a $\pi$ phase shift for light with a wavelength of 780~nm. 

\section{Amplitude Shaping}
\begin{figure*}
\includegraphics{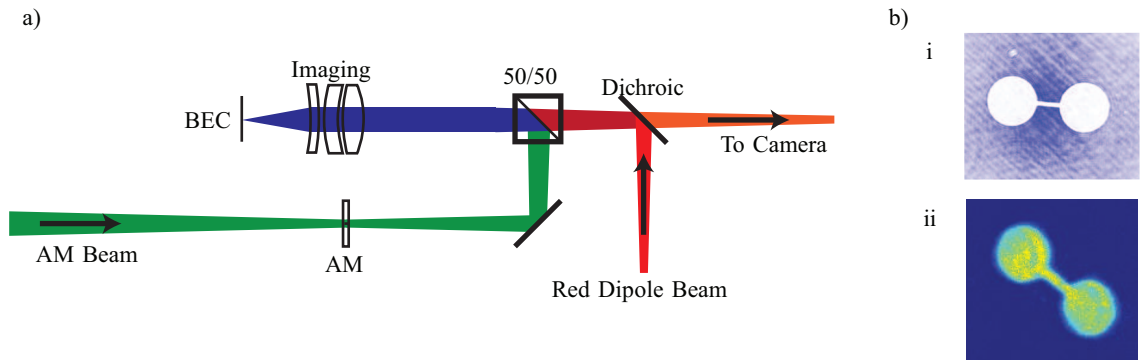}
\caption{a) Schematic of the imaging setup used for the AM system. This setup combines the AM dipole trap beam with a red-detuned (1064~nm) dipole trap beam and counter-propagating imaging beam (589~nm) using a 50/50 beamsplitter. These two beams are separated using a dichroic plate beamsplitter. All three of these beams are imaged through a main imaging stack consisting of an air-spaced triplet lens stack on the higher NA atom side, and and a cemented doublet on the lower NA optical table side. The mask is placed on a 3-axis translation stage to allow precise positioning. b) i) A typical CCD camera image of the output plane of a AM optical setup. This example is from a simple, 1:1 system using a single spherical lens. ii) An in-situ absorption image of a sodium BEC trapped in this potential. \label{fig:DIM}}%
\end{figure*}

Amplitude shaping, the second method used for creating arbitrary 2D intensity patterns, is considerably more lossy as it involves an opaque amplitude mask (AM) to block part of the beam.  The setup, however, is simpler than the phase-shaping setup and the total cost is considerably lower.  All that is required is an amplitude mask and a series of lenses to relay the shadow of the mask onto the atoms as sketched in Figure~\ref{fig:DIM}. For the blue-detuned light used in this demonstration, the mask blocked light associated with the trapping regions while transmitting the light responsible for the dipole potential.  As with phase shaping, amplitude shaping is also easily integrated into commonly used imaging systems for cold atoms. In our demonstration, we combined the imaging for the amplitude shaping with that used for absorption imaging diagnostics. This entire optical setup, as well as a demonstration of atoms trapped with this technique, is shown in Figure~\ref{fig:DIM}. 

Loading the trap required some care, especially with blue detuning, as the optical dipole trap was loaded from a larger magnetic trap. The Gaussian envelope tends to push atoms initially on the fringes of the magnetic trap farther from the desired trapping region. Thus, to improve transfer, we first transfer the atoms into a 1064~nm dipole trap. The non-adiabatic change of trap geometry both compresses the atoms to a convenient size for transfer to the blue-detuned trap, and increases the phase-space density, condensing the atoms into a BEC. The atoms are then adiabatically transferred to the AM dipole trap. 

\subsection{Contruction of Intensity Masks}
For our experiments designed to use the AM, 2D potential landscapes had to be created that extended over a length of up to approximately 150~$\mu$m, with minimum structure sizes down to approximately 10~$\mu$m. The imaging system that was used for absorption imaging of the atoms as well as to localize the two additional dipole traps, provided a reduction of 2.9 between the plane of the atoms and the image plane on the optical table. The dimensions of the necessary masks are thus approximately 450~$\mu$m with features around 30~$\mu$m. While this is incredibly difficult to achieve with ordinary machining techniques, it is well within the realm of possibility of even very basic photolithographic techniques used to fabricate microelectronic circuits. With the wide availability of fabrication labs at many universities, as well as private facilities, the necessary masks can be created quickly with minimal training. Special care needs to be taken, however, beyond standard procedure in these labs, as optical quality is generally not a concern in microelectronic applications.

The process to fabricate a mask starts with a standard anti-reflective (AR) coated optical window, with the desired size and coating necessary for the particular species being trapped. For this experiment, 1/2" windows were used with a broadband AR coating (350--700~nm). The window was first coated on one side with a $\sim$200~nm thick layer of chrome using thermal vapor deposition, which was sufficient to produce an experimentally verifiable optical density greater than 7. The thickness was constantly monitored during deposition with a calibrated quartz oscillator that was also exposed to the chrome. A special rig to hold the windows in the vacuum chamber was necessary to avoid scratching the back side of the window, and standard optical cleaning procedures were carried out before deposition.

The chrome coated windows were then coated with a layer of photoresist (Shipley 1813), which is spun on using standard methods for silicon wafers. Again, proper tools must be constructed and used to ensure the back side of the window is not scratched during photoresist spinning, or the subsequent baking to cure the photoresist. Since standard procedures in the lab are designed for thin (0.5~mm) silicon wafers, the baking time was adjusted to accommodate the relatively thick (3~mm) window. For initial tests, the time was approximately doubled to two minutes. The roughness of the edges of the final masks produced was on the scale of 5~$\mu$m. It is expected that this is due to improper baking since uneven adhesion of the photoresist layer can cause these effects. Given the image magnification and diffraction limit of the experimental setup used however, this was more than sufficient for our purposes. It is expected that further fine-tuning of the baking time, which would improve photoresist adhesion, would result in smoother masks if necessary for future experiments. 

Once the photoresist is exposed to ultraviolet (UV) light, the polymer breaks down, and can be removed by a weak developer solution that leaves unexposed photoresist intact. In order to carry this process out selectively, and create our desired pattern, a commercially produced mask is used with the same dimensions as our final product. These can come in many forms, but those used in this experiment are created on a Mylar film with an opaque resin. These are easily produced from source files in a standard Gerber format. This mask is sufficient for exposing the photoresist, but lacks the AR coating desired, and optical density necessary for the experimental intensity mask.

This commercial mask is placed over the chrome and photoresist coated window in an aparatus known as a mask aligner. This machine combines a vacuum system to hold the window and mask in place, a precision translation and rotation stage to align the two, a microscope to inspect alignment, and a timed UV exposure system. Again, another custom holder needs to be made to hold the window without scratching the back side, and to hold the window lower than the typically thinner silicon wafers used. Once exposed, the mask is removed and the window is again baked, with adjusted baking time, to finish curing the photoresist. It is then rinsed in the developer solution for approximately 40~seconds to remove the exposed photoresist. The result can then be examined under a microscope to verify that the desired pattern has been transferred properly. If not, the remaining photoresist can be stripped with acetone, and the photoresist, exposure, developing procedure can be attempted again. Once the proper transfer has been verified, the window is rinsed in a chrome etchant solution for approximately 2~minutes, or until the chrome in the exposed regions has been completely etched off of the window. The chrome etchant used in this experiment had no noticeable effect on the AR coating of the window. Any remaining photoresist can then be removed, and the finished chrome intensity mask is left on the optical window, and can be placed in the experimental setup at the object plane of the imaging system.

\section{Advantages}
\subsection{Phase-Contrast Shaping}
There are two main advantage of phase shaping over amplitude shaping. First, the theoretical efficiency of the former, defined as the fraction of total input power contained in the desired output pattern, can be larger be more than a factor of 2, depending on the exact pattern used. Experimentally, however, this is not always the case. For our demonstration\cite{Lee:2013}, the phase spot used was extremely small, and ideally should have been larger, allowing diffraction limited performance down to this size in the Fourier plane of the optical setup. This was not possible with the spherical optics used, contributing to the reduced efficiency. More specialized aspherical lenses or complex lens systems could be used in future generations of the setup to minimize this source of loss.

The second, and more important, advantage is the flexibility this method provides when combined with an SLM. For the experiment we performed\cite{Lee:2013}, the SLM was computer controlled, and is interfaced to the computer as an external monitor. With this setup, changing the pattern is just a matter of changing what is on that monitor. While simulation are accurate enough to provide a good starting point for pattern designs, the pattern on the SLM can be fine tuned to increase efficiency once the optical setup is complete. In addition to fine tuning individual pattern parameters, patterns can be completely changed in a matter of seconds as well, providing quick adjustment to experimental parameters for certain experiments. In addition, it may also be possible to change the PM on a time scale commensurate with degenerate gas dynamics.

\subsection{Amplitude Shaping}
There are two main advantages for the AM approach as well. First, it is extremely simple to implement. All that is strictly necessary for this setup to work is a light source for the dipole trap, and a lens to image the plane of the mask to the plane of the atoms. The spatial resolution of the final trap will be limited only by the resolution of the imaging system. Because many cold atom experiments already have a need for a high resolution imaging systems to observe the atoms, this same system can often be used for both purposes if designed with this in mind. The second advantage of this technique is the cost of implementation. With the simplicity of the optical setup, time can be saved in putting the mask system into place as well.

\section{Limitations}
\subsection{Phase Shaping}
The main limitation of the GPC technique, when implemented with an SLM, is the cost of SLM. SLMs are also generally designed for limited ranges of wavelengths, limiting the flexibility of the system. In addition, the technical details of how the SLM works, namely the necessary input polarization and the inter-pixel dead-space, can affect the resulting potential that the SLM produces. In order for the SLM to change only the phase of the light incident upon it, the light must be linearly polarized in a particular direction. For reflective SLMs, which are preferable due to reduced dead-space as discussed below, this means that one of two things can be done to separate the input light from the output. Either the light incident on the SLM must be at a slight angle, or a 50/50 beamsplitter can be used if the light must come in normally. The first option can be problematic if the depth of field of the imaging sytem in the GPC setup is smaller than the longitudinal displacement of the tilted SLM plane. If the second option is used to eliminate this problem, the total power is cut down to 25\% of the input. Combined with the relatively low damage threshold for SLMs, this can be a problem for far-detuned traps. A similar photolithographic process as for the AM could be used to etch a phase mask. While this would negate all of the problems mentioned above, it would also result in a static mask without the versatility of the SLM created masks.

Before ending this section, we return to the dead-space mentioned above. In any pixelated structure, like an SLM, there will necessarily be some space between pixels that cannot be used. In transmissive SLMs, this space contains the wires that address each pixel, and is blocks the light. While reflective SLMs can improve significantly on this, there is still some unusable space between pixels. The reflective SLM used in this experiment has a $>$90\% fill factor, so this dead-space is quite small, but is still enough to cause a noticeable effect in the output. In our phase shaping scheme, this shows up as a slight intensity drop in the space between pixels as can be seen in Figure~\ref{fig:GPC}. Using a blue detuned trap minimizes the problems this can cause, since the atoms are trapped in areas with no light. Additionally, some newer, reflective coated, liquid crystal on silicon (LCoS) SLMs claim to have near 100\% fill-factors, however we have not tested these devices to date. Like traditional SLMs, the LCoS SLMs can be used for both phase and amplitide shaping\cite{Frumker:2007b, Vaugha:2005, Frumker:2007a}.

\subsection{Amplitude Shaping}
The first limitation of amplitude shaping is that it is much less flexible when it comes to changing patterns when an etched mask is used. When a different, existing pattern is desired, the mask must be physically replaced and realigned. To aid in this process in our setup, the mask was placed on a three-axis translation stage, and realignment times for new masks were reduced to about 15~minutes. If a completely new mask is desired, the whole fabrication process must be redone, taking up to a few days. As mentioned above, and SLM could also be used to create the AM, however we have not tested this method.

The other limitation that we encountered was due to absorption of light by the chrome making up the mask. The reflectance of chrome at 532~nm is only approximately 60\%, meaning that 40\% of the light incident on the chrome was being absorbed. Due to multiple 50/50 beam splitters after the mask for combining all the beams going through the imaging system, up to 500~mW of light was incident on the mask in a beam with $1/e^2$ radius of about 200~$\mu$m, giving intensities of greater than 7~MW/$\mathrm{cm}^2$. Depending on the exact geometry and smallest feature size on a particular mask, both of which will affect the rate of heat dissipation greatly, this was enough to heat up the chrome to the point of delaminating from the window. Future generations of this technique will implement other higher reflectance metals such as silver to limit this deficiency. Additionally, it was noticed that the damage threshold was significantly increased by illuminating the mask from the back (through the window), presumable due to surface contamination contributing to laser absorption. This is also the preferred side to illuminate to minimize imaging aberrations.

\section{Conclusions}
We have shown two new methods for creating arbitrary 2D ODP for use in atomic physics experiments. The generalized phase-contrast method, while being more expensive and being limited in wavelength range and usable power, has more flexibility in being able to quickly adjust trapping parameters, and design new trapping geometries. With the direct intensity masking method it is much easier to create and design patterns, while being more flexible for fast changes. The flexibility of the DIM method can be increased significantly though, with the addition of an AOD scanned Gaussian beam. With the main trap being a static mask potential, and the AOD beam limited to relatively small perturbations to this, the higher heating rates seen in AOD traps is not observed, and the advantage of dynamic traps is still available.

\begin{acknowledgments}
The authors acknowledge technical support from the staff of the FabLab at the University of Maryland, as well as G. Campbell, S. Eckel, F. Jendrzejewski and A. Kumar on the amplitude shaping, as well as C. Alonzo, N. Chattrapiban, R. van Fleet and B. McIlvain on the phase-contrast shaping.  This work was done in partial fulfilment of the PhD thesis of JGL and funded in part by the Physical Frontier Center at the Joint Quantum Institute (PHYS 0822671).
\end{acknowledgments}

\end{document}